\documentstyle[prl,multicol,aps,epsfig]{revtex}
\begin{document}

\title{Non-analytic dependence of the transition
 temperature of the homogeneous dilute Bose gas on scattering length}
\author{Markus Holzmann and Gordon Baym}
\address{University of Illinois at Urbana-Champaign,
1110 W. Green St., Urbana, Il 61801, USA}
\author{Jean-Paul Blaizot}
\address{CEA-Saclay, Service de Physique Th\'{e}orique,
91191 Gif-sur-Yvette, Cedex, France}
\author{and}
\author{Franck Lalo\"e}
\address{Laboratoire
Kastler Brossel,
Ecole Normale Sup{\'{e}}rieure,
24, rue Lhomond, 75231 Paris Cedex 05, France}
\maketitle

    \begin{abstract} We show that the shift in the transition temperature of
the dilute homogeneous Bose gas is non-analytic in the scattering amplitude,
$a$.  The first correction beyond the positive linear shift in $a$ is negative
and of order $a^2\ln a$.  This non-universal non-analytic structure indicates
how the discrepancies between numerical calculations at finite $a$ can be
reconciled with calculations of the limit $a \to0$, since the linearity is
apparent only for anomalously small $a$.
\end{abstract}

\begin{multicols}{2}
\narrowtext

    The shift of the critical temperature of the dilute homogeneous Bose gas,
$T_c$, with interactions has had a long and controversial history
\cite{lee,stoof,stoof2,peter,ursell99,linres,gordon1,gordon2,N2,svistonov,arnold,bigbec}
(and further references therein).  The correction to the ideal gas critical
temperature, $T_c^0$, is now established theoretically \cite{gordon1,gordon2}
(see also \cite{linres}) to be linear in the scattering length $a$ in leading
order
\begin{equation}
\frac{\Delta T_c}{T_c^0}=\frac{T_c - T_c^0}{T_c^0} = c \, a n^{1/3},
\quad a n^{1/3} \to 0,
\label{one}
\end{equation}
where $n$ is the particle density.  The variation of $T_c$ is determined
primary by large distance, or small momenta, contributions.  Due to infrared
divergencies around the transition point, pertubation theory is not
applicable, and the coefficent $c$ cannot be obtained by simple perturbative
techniques.  Except in special cases, e.g., a large number of internal
degrees of freedom, the constant $c$ must be evaluated numerically.

    However, even numerical calculations have not settled this issue.  Whereas
Ref.~\cite{linres} provides $c=2.33 \pm 0.25$ from a direct calculation of the
%%%%%%%%%%% 2.33 %%%%%%%%%%%%%%%
coefficient first taking the limit $a n^{1/3} \to 0$, Ref.~\cite{peter}
predicts $c=0.34 \pm 0.06$ after numerical extrapolation of the calculation to
the limit $a \to 0$.  In this paper we show that the difference
in these two results is attributable to
 a non-analytic structure, $\sim a^2 \ln a$, of
the transition temperature in $a$.  This correction, which is negative, does
not introduce a new length scale beyond $n^{-1/3}$, but because of its
logarithmic character it gives rise to a strong dependence on $a$ even in the
very dilute limit, $a n^{1/3}\to 0$.  Very recently, Refs.
\cite{svistonov,arnold} obtain $c \simeq 1.3$ from a classical $\phi^4$ model
on a lattice extrapolated to the continuum, a result qualitatively consistent
with the larger value found in Ref.~\cite{linres}.  In these calculations the
logarithmic terms enter as corrections to the continuum limit.

    Equation (\ref{one}) is only the beginning of an asymptotic expansion,
as one might suspect, since for $a<0$ the system is unstable.  Inclusion of
the $a^2 \ln a$ term, when one extrapolates numerical data from finite $a$
values to the limit $a \to 0$, provides a first resolution of the apparent
discrepancies between numerical calculations done at finite $a$ \cite{peter}
and those valid for $a \to 0$ \cite{linres,svistonov,arnold}.  To obtain a
quantitative estimate, we explicitly calculate the logarithmic correction in a
model with $N$ internal degrees of freedom, to leading order in $1/N$.  The
result suggests that the linear increase
%%%%%%%%%%%%%%%%%%%%%%%%%%%%%%%%%%%%%%%%%%%
 of $\Delta T_c$ at small but finite $a$ is noticeably
%%%%%%%%%%%%%%%%%%%%%%%%%%%%%%%%%%%%%%%%%%%
suppressed for the physical case of $N=2$.

    We consider a uniform system of bosons of mass $m$ at temperature $T$, and
assume that the two-body interaction can be described by the $s$-wave
scattering length $a$.  Above the critical temperature, the density $n$ is
given in terms of a sum over Mastubara frequencies, $z_{\nu}=2 \pi i \nu T$
($\nu= \pm 1, \pm2,\ldots$), of the single particle Green's function $G(k,z)$:
\begin{equation}
n = - T \sum_{\nu} \int \frac{d^3k}{(2\pi)^3} G(k,z_{\nu}),
\label{two}
\end{equation}
where ($\hbar=k_B=1$)
\begin{equation}
G^{-1}(k,z)=z +\mu - \frac{k^2}{2m} -\Sigma(k,z),
\end{equation}
and $\mu$ is the chemical potential; the condition $\mu = \Sigma(0,0)$
determines the Bose-Einstein condensation point.

    The shift of the critical temperature at fixed density is more
conveniently calculated in terms of the shift, $\Delta
n_c=n_c(a,T_c)-n_c(0,T_c)$, in the critical density at fixed $T$; the two
shifts are related, to the orders of interest (less than $a^2$), by
$\Delta T_c/T_c = -(2/3) \Delta n_c/n_c$.  As shown in \cite{gordon1}, the
leading linear shift $\Delta T_c^{(1)}$ is given solely by the zero
Matsubara frequency term:
\begin{eqnarray}
\frac{\Delta T_c^{(1)}}{T_c^0}
& = &
\frac{2 T_c^0}{3 n} \int \frac{d^3k}{(2\pi)^3} \left[ G(k,0)-G_0(k,0) \right]
\\
&= &
\frac{4 \lambda}{3\pi \zeta(3/2)} \int_0^{\infty} \,d k \,
\frac{U(k)}{k^2 +U(k)},
\label{lo}
\end{eqnarray}
where $\lambda=(2 \pi/mT)^{1/2}$ is the thermal wavelength,
$\zeta(3/2)=2.612\ldots$, $G_0$ is the Green's function of the ideal Bose gas
at $T_c^0$, and
\begin{equation}
U(k)=2m (\Sigma(k,0) - \Sigma(0,0)).
\end{equation}
At the critical temperature, $U(k)$ can be calculated to order
$a^2/\lambda^4$ by considering only the $\nu=0$ sector, which corresponds to a
classical field theory.  At the transition, $U(k)$ has the scaling structure,
\begin{equation}
U(k)= \frac{a^2}{\lambda^4} \sigma( k \lambda^2/a),
\end{equation}
from which the linearity of $\Delta T_c^{(1)}$ in $a$ follows
\cite{gordon1,gordon2}.
If $U(k)$ is calculated by classical field theory, $\Delta T_c^{(1)}$ is
strictly linear in $a$.

    The next-to-leading order corrections, $\Delta T_c^{(2)}$, arise in terms
with non-zero Matsubara frequencies, both explicitly [Eq.~(\ref{two})]
 and in internal loops, in
the calculation of $U(k)$.  As we show below, the internal loop corrections
begin at order $a^2$ and $a^3\ln a$; the $a^2\ln a$ terms may be extracted
from
\begin{equation}
\frac{\Delta T_c^{(2)}}{T_c}
 =
\frac{2 T_c^0}{3 n} \sum_{\nu \ne 0} \int
\frac{d^3k}{(2\pi)^3} \left[ G(k,z_{\nu})-G_0(k,z_{\nu}) \right].
\end{equation}
Since the infrared behavior is regular for $\nu\ne 0$, we expand the
denominator of $G$ to first order in $\Sigma(k,z_{\nu})-\mu$, and write
\begin{eqnarray}
\frac{\Delta T_c^{(2)}}{T_c}& &\nonumber\\
 \simeq  & &
 \frac{2 T_c^0}{3 n} \sum_{\nu \ne 0} \int
\frac{d^3k}{(2\pi)^3}
\frac{\Sigma(k,z_\nu)-\Sigma(k,0) + U(k)/2m}{(z_\nu-k^2/2m)^2}.\nonumber\\
\label{nonzero}
\end{eqnarray}
From pertubation theory we know the functional form of $U(k)$ outside
the critical region in $k$,
\begin{equation}
U(k) \propto  \frac{a^2}{\lambda^4} \ln \frac{ k \lambda^2}{a},
\quad k \gg \frac{a}{\lambda^2};
\label{log}
\end{equation}
this logarithmic behavior is valid to all orders of pertubation theory, as
one can verify by power counting.  Since the dominant contribution to
the integral in Eq.~(\ref{nonzero}) comes from momenta $k \sim \lambda^{-1}$,
the 
ultraviolet behavior of $U(k)$ generates a logarithmic shift 
in the critical temperature:
\begin{equation}
\frac{\Delta T_c^{(2)}}{T_c}
\propto \frac{a^2}{\lambda^2} \ln \frac{a}{\lambda}.
\label{nlo}
\end{equation}
Since $\Sigma(k,z_\nu)-\Sigma(k,0)$ tends to zero for large momenta $k$,
the contribution of this term in Eq.~(\ref{nonzero}) remains of order
$a^2/\lambda^2$.  Thus the next-to-leading order to the critical temperature
shift is proportional to $a^2\ln a$ and is always {\it negative} for small
$a$.

    In order to estimate the shift quantitatively, we calculate it in the
large $N$ model.  The $\nu=0$ sector, equivalent to a classical $\phi^4$ field
theory in three spatial dimensions, is described by the action \cite{zinn},
\begin{eqnarray}
S\{ \phi(r) \}
& = & \int d^3 r \left\{ \sum_{i=1}^{N} \frac12 [\nabla \phi_i(r)]^2
\right. \nonumber \\
&^{}& \left.
- m \mu T \sum_i \phi_i^2(r) + \frac{u}{4!} \left[ \sum_i \phi_i^2(r)
\right]^2 \right\},
\end{eqnarray}
with $u=96 \pi^2 a/\lambda^2$.  The classical field theory suffers from
ultraviolet divergencies, which can be regularized by introducing a large
momentum cutoff $\Lambda$.  As shown in Refs.  \cite{gordon1,gordon2}, the
leading order corrections to the critical density are dominated by long
distance properties and $U(k)$ is independent of the cutoff.  Therefore one
can derive $U(k)$ with a fixed cutoff $\Lambda$ in the action, take the limit
$\Lambda \to \infty$, and determine the corrections to the critical density
from Eqs.  (\ref{lo}) and (\ref{nonzero}).

    Instead of this procedure we will obtain the next-to-leading order
corrections in an independent way, which has the advantage of making contact
with the numerical $\phi^4$ lattice calculations.  Starting from the finite
temperature quantum field action, one can derive the effective action of the
classical field theory by integrating pertubatively over the non-zero
frequency quantum modes, $\nu \ne 0$, which provides a large momentum cutoff
$\Lambda \sim \sqrt{m T} \sim 1/\lambda$ and renormalized effective
coefficients of the Euclidean action \cite{zinn}.  Following \cite{gordon2},
the corrections to the transition temperature are given in this effective
field theory by
\begin{equation}
\frac{\Delta T_c}{T_c}
= \frac{4 \lambda}{3 \pi \zeta(3/2)} \int_0^{\Lambda} dk
 \frac{U_{\Lambda}(k)}{k^2 + U_{\Lambda}(k)},
\end{equation}
where the subscript $\Lambda$ indicates the explicit dependence on the
ultraviolet cutoff which incorporates the leading effects of non-zero
Matsubara frequencies.

    In the large $N$ limit $U_\Lambda(k)$ is given in terms of the
particle-hole bubble, $B(q)$, by
\begin{eqnarray}
U_{\Lambda}(k)
= - \frac{Nu^2}{18} \int^{\Lambda}_0&& \frac{d^3q}{(2\pi)^3}
\frac{B(q)}{1+N u B(q)/6} \nonumber \\
&&\times \left[ \frac{1}{(k-q)^2}-\frac{1}{q^2} \right];
\end{eqnarray}
to leading order in $1/N$ in three dimensions \cite{zinn},
\begin{equation}
 B(q)= \frac{1}{8 q} - \frac{6}{Ng^* \Lambda} + {\cal O}(\Lambda^{-2})
\label{bubb}
\end{equation}
where $g^*=48 \pi^2/N$.  Following Ref.  \cite{gordon2}, we obtain the
critical temperature shift
\begin{equation}
\frac{\Delta T_c}{T_c}=\frac{4 \lambda}{3\pi\zeta(3/2)}
\int_0^{\Lambda} dk \frac{U_{\Lambda}(k)}{k^2}.
\end{equation}
As we see from the $\Lambda$-dependence of the bubble, Eq.~(\ref{bubb}), the
$\Lambda$-dependence of $U_{\Lambda}(k)$ gives rise to higher order
corrections in $\Lambda^{-1}$.  These terms arise effectively from the
internal non-zero Matsubara frequencies, and lead to corrections
$\sim\Lambda^{-2} \ln \Lambda$ in $\Delta T_c/T_c$, which we can neglect.
Thus
\begin{eqnarray}
\frac{\Delta T_c}{T_c} & = &-\frac{64 \lambda}{3\pi\zeta(3/2)}
\frac{a}{\lambda}
\int_0^{\Lambda\tau} dk
\int_0^{\Lambda\tau/k} dx
\frac{1}{k(xk+1)}
\nonumber \\
&&  \times \left[
\frac{x}{2} \log \frac{|1+x|}{|1-x|} -1 \right],
\end{eqnarray}
where $\tau=(N u/48)^{-1}$.  To obtain the leading order corrections we
differentiate with respect to $\Lambda \tau$, which allows us to isolate the
contributions around $\Lambda \tau \rightarrow \infty$.  Integrating back we
find
\begin{equation}
\frac{\Delta T_c}{T_c}= \frac{8\pi}{3\zeta(3/2)}\frac{a}{\lambda}
 \left\{
  1 + 16 N \frac{a}{\lambda^2 \Lambda}
   \ln \frac{N a}{\lambda^2 \Lambda}
+ {\cal O}\left( \frac{N a}{\lambda^2 \Lambda} \right)
\right\}.
\label{tcres}
\end{equation}
The integration constant is given by the large $N$ result for the linear
shift, Ref.~\cite{gordon2}.  In contrast to the linear corrections in $a$,
the next-order terms depend on the short length scale properties of the
system, modelled by the cutoff $\Lambda$, and therefore are not universal.

    To estimate the influence of the logarithmic terms we take $N=2$ and
$\Lambda \lambda =(2 \pi)^{1/2}$ in Eq.~(\ref{tcres}).  The corrections to the
linear behavior of $T_c$ found in this way are precisely those one finds for
$N=2$ by including the particle-hole bubble sum in $U(k)$ . The resulting
dependence of the transition temperature on $a$ is shown in Fig. 1. For a gas
parameter $na^3 \sim 10^{-6}$, corresponding to the experimental region of
Bose-Einstein condensation in atomic gases \cite{BEC} and liquid $^4$He in
Vycor \cite{reppy}, and to the lowest density Monte Carlo data of
Ref.~\cite{peter}, the nonlinear corrections depress the linear shift by $\sim
50\%$; instead of the $a\to 0$ result $c \simeq 2.33$ in Eq.~(\ref{one}), one
obtains a coefficient $c \sim 1.2$ for $an^{1/3} \sim 10^{-2}$.  Even if the
extrapolation from the large $N$ expansion to $N=2$ is unjustified, this
calculation suggests that the logarithmic terms play an important role for
present numerical and experimental parameters.  The noticeable depression of
$\Delta T_c$ in this parameter regime is also confirmed by self-consistent
numerical model calculations \cite{bigbec}.

    A qualitatively similar strong dependence on $a$ is found in the
renormalization group calculations of Ref.~\cite{stoof2}, which derives an
$a\ln a$ correction \cite{stoof3}.  Such a result would follow from
Eq.~(\ref{lo}) were $U(k)$ to be linear in $k$ up to an ultraviolet cutoff.
However, in the regime $a/\lambda^2 \ll \Lambda$, corresponding to the dilute
limit, $a/\lambda\ll 1$, $U(k)$ is given by the perturbative result
(\ref{log}) for momenta in the region $a/\lambda^2 \ll k \ll \Lambda$ rather
than being linear in $k$.

\begin{figure}
\begin{center}
\epsfig{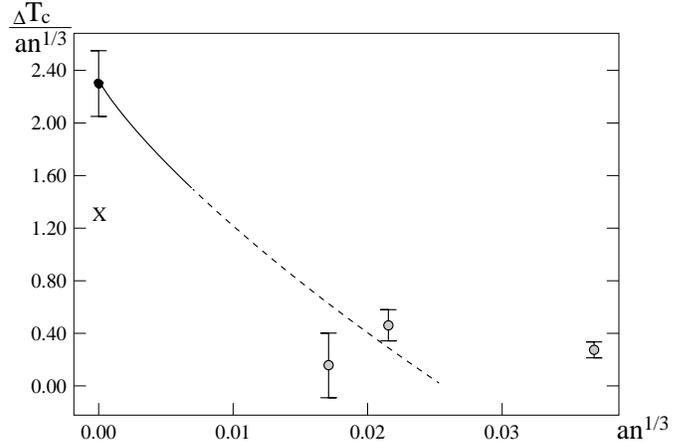}
\end{center}
\caption{
Dependence of the transition temperature of a dilute homogeneous
Bose gas on scattering length, from Eq.~(\ref{tcres}) for $N=2$.  The falloff
of $\Delta T_c/an^{1/3}$ with increasing $an^{1/3}$ arises from the
non-universal next-to-leading order logarithmic corrections.  In the regime
where the curve is represented by a dashed line, higher order corrections
begin to become important.  The dark circle is the calculation of
Ref.~[6]; the points shown as open circles are the lowest density
data from the numerical calculations of Ref.~[4]. The data point
shown by the cross indicates the numerical results of
Refs.~[10,11]
for a lattice $\phi^4$ theory extrapolated to the continuum.
}
\end{figure}

    The next-to-leading order corrections are important for classical $\phi^4$
calculations as well \cite{svistonov,arnold}; in the universal region where
the influence of the ultraviolet cutoff $\Lambda$ is unimportant ($\Lambda \to
\infty$), classical field theory provides perfect scaling, implying a linear
shift of $T_c$ for {\em all} $a$.  Variations in $a$ of $\Delta T_c$ in
\cite{svistonov} are due to non-universal corrections and are sensitive
to the details of the scheme used to regularize the classical $\phi^4$ theory
\cite{arnold2}.  Extrapolation to the universal small coupling region, $a/\lambda^2\Lambda
\to 0$, together with finite size scaling to the thermodynamic limit allows
one not only to extract the coefficient $c$ of the linear shift in $T_c$, but
should also provide the magnitude of the non-universal $a^2 \ln a$ corrections
for the physical case $N=2$.

We dedicate this article to our recently deceased friend,
Dominique Vautherin, with whom we had many stimulating discussions
on this subject.
We are grateful to Henk Stoof for stimulating discussions both during his
Workshop on Bose-Einstein condensation at the Lorentz Center in Leiden, and at
the Ecole Normale Sup\'erieure.
We also thank Peter Arnold, Nikolay Prokof'ev, and Boris Svistunov
for helpful comments on their lattice calculations.  Author GB would like to
thank the ENS for its hospitality in the course of this work.  This research
was facilitated by the Cooperative Agreement between the University of
Illinois at Urbana-Champaign and the Centre National de la Recherche
Scientifique, and supported in part by the NASA Microgravity Research
Division, Fundamental Physics Program and by National Science Foundation Grant
PHY98-00978.  Laboratoire Kastler Brossel de l'Ecole Normale Sup\'erieure is
{\it UMR 8552 du CNRS} and {\it associ\'e \`a l'Universit\'e Pierre et Marie
Curie}.

%\bibliographystyle{prsty}
%\bibliography{refbec,refman}

\end{multicols}
\end{document}